# Anforderungen an die Qualitätssicherung von KI-Modellen für die Lungenkrebs-Früherkennung

*Horst K. Hahn • Matthias S. May • Volker Dicken • Michael Walz • Rainer Eßeling • Bianca Lassen-Schmidt • Robert Rischen • Jens Vogel-Claussen • Konstantin Nikolaou • Jörg Barkhausen*

**Institutionen:** Fraunhofer MEVIS, Institut für Digitale Medizin, Bremen (H.K.H., V.D., B.L.S.), AG Digitale Medizin, FB 3, Universität Bremen (H.K.H.), Radiologisches Institut, Universitätsklinikum Erlangen, Friedrich-Alexander-Universität Erlangen-Nürnberg (M.S.M.), Imaging Science Institute, Universitätsklinikum Erlangen (M.S.M.), Ärztliche Stelle für Qualitätssicherung in der Radiologie, Nuklearmedizin und Strahlentherapie Hessen TÜV SÜD Life Service GmbH, Frankfurt (M.W.), Universitätsklinikum Münster, Klinik für Radiologie (R.E., R.R.), Medizinische Hochschule Hannover, Institut für Diagnostische und Interventionelle Radiologie (J.V.C.), Universitätsklinikum Tübingen, Abteilung für Diagnostische und Interventionelle Radiologie (K.N.) und Universität zu Lübeck, Institut für Radiologie und Nuklearmedizin (J.B.). **Korrespondenz an:** H.K.H. (horst@uni-bremen.de) & J.B. (joerg.barkhausen@uksh.de)

## Zusammenfassung

Lungenkrebs ist die zweithäufigste Krebserkrankung und die häufigste Krebstodesursache weltweit. Die Überlebenschancen hängen entscheidend vom Tumorstadium zum Zeitpunkt der Diagnose ab und zahlreiche Studien haben nachgewiesen, dass bei Hochrisikopatientinnen und -patienten die Lungenkrebsfrüherkennung mittels Niedrigdosiscomputertomographie (LDCT) die Lungenkrebsmortalität erheblich senken kann. Nachdem in Deutschland seit Sommer 2024 eine Rechtsverordnung des Bundesministeriums für Umwelt, Naturschutz, nukleare Sicherheit und Verbraucherschutz (BMUV) die Anwendung der Computertomographie zur Lungenkrebsfrüherkennung nach Genehmigung durch die Aufsichtsbehörde erlaubt, läuft aktuell das Verfahren des Gemeinsamen Bundesausschusses (G-BA) zur Aufnahme der Methode in den Leistungsumfang der Gesetzlichen Krankenversicherung.

Viele Studien belegen das Potenzial künstlicher Intelligenz (KI), die Genauigkeit der Detektion, Volumenbestimmung und Charakterisierung von Lungenrundherden (Abb. 1) zu verbessern und die radiologische Befundungszeit zu verkürzen. Daher hat das BMUV in seiner Rechtsverordnung in den Anforderungen für

> ### *Kernaussagen des Positionspapiers*
>
> - Einheitliche und transparente Qualitätsstandards für KI-Anwendungen stellen eine wesentliche Voraussetzung für eine hochwertige radiologische Untersuchung dar, z.B., um zu verhindern, dass suboptimale Softwaresysteme zu Überdiagnosen und unnötigen Behandlungen führen oder Läsionen übersehen werden.
> - Radiologische Institutionen erhalten durch eine objektive Prüfung Klarheit bei KI-Auswahl, Ausschreibungen und Kaufentscheidungen.
> - Eine unabhängige Prüfung sorgt für Akzeptanz in Genehmigungen, Leitlinien oder Positionspapieren, verringert das Risiko, dass Hersteller ausschließlich auf einzelne, ausgewählte Datensätze hin optimieren, und stärkt insgesamt das Vertrauen in die Technologie.
> - Ein Datensatz, in dem klinisch relevante Lungenkarzinome sicher belegt sind, bietet perspektivisch die Möglichkeit einer objektiven, auch diagnosebezogenen Bewertung der KI-Leistung.
> - Physische Phantome und simulierte digitale Läsionsphantome tragen als Ergänzung eines repräsentativen in-vivo Datensatzes wesentlich dazu bei, die Erkennung und volumetrische Genauigkeit belastbar zu überprüfen.
> - Die technischen und finanziellen Ressourcen für den Aufbau und die Pflege von Referenzdatensätzen sowie die Durchführung von Prüfprozessen sind eine nachhaltige Investition in ein hochwertiges und kosteneffizientes Screeningverfahren.
> - Verzögerungen bei der Markteinführung hochwertiger und zuverlässiger KI-Lösungen können durch frühzeitiges, proaktives Handeln minimiert werden.

*Summary*

**Requirements for Quality Assurance of AI Models for Early Detection of Lung Cancer**

*Lung cancer is the second most common type of cancer and the leading cause of cancer-related deaths worldwide. The chances of survival largely depend on the tumor stage at the time of diagnosis. Numerous studies have shown that early detection of lung cancer using low-dose computed tomography can significantly reduce lung cancer mortality in high-risk patients. Since summer 2024, a legal regulation issued by the German Federal Ministry for the Environment (BMUV) has permitted the use of computed tomography for lung cancer screening following approval from the regulatory authority. Currently, the German Federal Joint Committee (G-BA) is in the process of deciding whether to include this method in the benefits covered by statutory health insurance.*

*Many studies highlight the potential of artificial intelligence (AI) to improve the accuracy of detecting, measuring, and characterizing pulmonary nodules while also reducing the time required for radiological assessment. Consequently, the BMUV has mandated the use of AI-based software for diagnostic support as part of its legal requirements for approval. However, the training data, functionality, and performance of available AI systems vary significantly, making it difficult for users to select appropriate software and for regulatory authorities to evaluate them during the approval process. While manufacturers are required to specify the intended application and provide supporting test statistics as part of the approval process, they are free to choose their training and test data. This lack of standardization means that there are currently few opportunities for direct comparison between different AI systems. At the same time, under the EU AI Act, the use of AI models for detecting, measuring, and characterizing pulmonary nodules requires consistent quality assurance by manufacturers.*

*To address this gap, this position paper proposes a systematic approach to quality assurance, centered around a validated reference dataset. This dataset should include both phantom data and real screening cases to enable precise measurement verification—particularly for volume and growth rate assessments—while also ensuring high relevance for lung cancer screening. Furthermore, the dataset should be continuously updated to reflect systematic changes in real-world data caused by demographic shifts or advancements in imaging technology. As a result, ongoing quality assurance of AI systems becomes essential.*

*Additionally, the paper highlights regulatory aspects. While the Medical Device Regulation (MDR) and the EU AI Act define basic requirements for approving AI systems, they inadequately address the unique challenges of self-learning algorithms and their updates during operation. A standardized and transparent quality assessment—based on factors such as sensitivity, specificity, and volumetric accuracy—allows for an objective and quantitative evaluation of the strengths and weaknesses of individual AI solutions. Establishing clear testing criteria and systematically using updated reference data form the foundation for comparable performance metrics, which can be incorporated into tenders, guidelines, and recommendations.*

*The German Radiological Society with its working groups on thoracic radiology, information technology (AGIT), and physics & technology (APT) aims to support this process, contributing to improved safety and quality in lung cancer screening. At the same time, the proposed framework can be applied to other radiological AI applications, further strengthening radiology's role as an innovative, quality-driven, and central medical discipline.*

eine Genehmigung den Einsatz einer entsprechenden Software zur Befundungsunterstützung vorgegeben. Allerdings variieren Trainingsbasis, Funktionalität und Performance verfügbarer KI-Systeme erheblich, wodurch die Auswahl geeigneter Software für die Anwender und die Bewertung, z. B. im Genehmigungsverfahren durch die Behörden, erschwert wird. Die Hersteller sind zwar verpflichtet, die beabsichtigte Anwendung und die dazugehörige Teststatistik im Zulassungsverfahren darzulegen, allerdings sind sie in der Wahl der Trainings- und Testdaten frei. Daher bestehen auf der einen Seite bislang kaum Vergleichsmöglichkeiten für Inverkehrbringer und Anwender. Die Verwendung von KI-Modellen zur Detektion, Volumenbestimmung und Charakterisierung von Lungenrundherden erfordert gemäß EU AI-Act auf der anderen Seite jedoch eine konsequente Qualitätssicherung durch die Inverkehrbringer.

Um diese Lücke zu schließen, schlagen wir mit diesem Positionspapier einen systematischen Ansatz zur Qualitätssicherung vor, bei dem ein valider Referenzdatensatz im Mittelpunkt steht. Dieser Datensatz sollte sowohl Phantom-Datensätze (vgl. Abb. 2) als auch reale, im Screening erfasste Fälle beinhalten, um einerseits exakte Überprüfungen der Messungen, insbesondere zur Bestimmung von Volumina und Wachstumsraten, zu ermöglichen und andererseits eine hohe Relevanz für das Screening darzustellen. Darüber hinaus sollte der Datensatz kontinuierlich weiterentwickelt werden, um systematische Veränderungen in den praktisch anfallenden Daten, die etwa durch demographische Veränderungen oder Fortschritte in der Bildgebung bedingt sind, adäquat abbilden zu können. Daraus ergibt sich auch die Notwendigkeit einer kontinuierlichen Qualitätssicherung der KI-Systeme.

Ergänzend rückt das Positionspapier regulatorische Aspekte in den Fokus. Die Medizinprodukte-Verordnung und der EU AI-Act definieren zwar grundlegende Anforderungen für die Zulassung von KI-Systemen, berücksichtigen jedoch die Besonderheiten lernender Algorithmen und deren Aktualisierung während des Betriebs nur unzureichend. Eine einheitliche und transparente Qualitätsbeurteilung – etwa anhand von Sensitivität, Spezifität und volumetrischer Genauigkeit – ermöglicht die objektive und quantitative Bewertung von Stärken und Schwächen einzelner Lösungen. Die Etablierung klarer Testkriterien und die strukturierte Nutzung aktualisierter Referenzdaten schaffen die Grundlage für vergleichbare Kennzahlen, welche in Ausschreibungen, Leitlinien oder Empfehlungen niedergelegt werden können.

Die Deutsche Röntgengesellschaft will diesen Prozess mit ihren AGs Thoraxradiologie, Informationstechnologie (AGIT) und Physik & Technik (APT) begleiten und damit einen wichtigen Beitrag zu Sicherheit und Qualität bei den Screening-Untersuchungen leisten. Gleichzeitig lässt sich das hier vorgestellte Konzept auf andere radiologische KI-Anwendungen übertragen, wodurch die Radiologie insgesamt in ihrer Rolle als innovatives, qualitätsorientiertes und zentrales medizinisches Fachgebiet gestärkt wird.

## 1. Aktueller Stand Lungenkrebs-Screening

Derzeit verfolgen mehrere Länder verschiedene Screening-Strategien, um Risikogruppen – insbesondere langjährige Raucherinnen und Raucher – flächendeckend zu erreichen (vgl. Kiraly et al., 2024; Pacheco et al., 2024). In Deutschland wurde im Sommer 2024 die Rechtsverordnung des BMUV zur Anwendung der Computertomographie zur Lungenkrebsfrüherkennung veröffentlicht (BMUV 2024). Seither ist die Durchführung der Untersuchung unter bestimmten Voraussetzungen nach Genehmigung erlaubt, und die G-BA-Entscheidung zur Aufnahme der Methode in den Leistungsumfang der gesetzlichen Krankenversicherung wird in diesem Jahr erwartet (vgl. Vogel-Claussen et al., 2024). In den USA wiederum werden bereits seit mehreren Jahren Screening-Programme umgesetzt, unter anderem auf Basis großer Studien wie dem National Lung Screening Trial (Berg et al., 2011). Derzeitige Leitlinien und Empfehlungen, insbesondere das Lung-RADS-System des American College of Radiology (aktualisierte Version von 2022), geben standardisierte Kriterien für die Befundung vor.

In diesem klinischen Kontext hat sich der Einsatz von KI-Modellen für die automatisierte Detektion, Volumenbestimmung und Charakterisierung von Lungenrundherden (vgl. Abb. 1) als vielversprechend erwiesen. So wurde etwa in der HANSE-Studie bereits ein entsprechendes Softwaresystem prospektiv erfolgreich implementiert (Vogel-Claussen et al., 2022). Die Algorithmen unterstützen Radiologinnen und Radiologen, indem sie auffällige Befunde markieren und zusätzliche Informationen – z. B. zu Größe, Wachstum, Morphologie oder Malignitätswahrscheinlichkeit – bereitstellen. Solche Modelle können dabei helfen, die Befundungszeit zu reduzieren und die diagnostische Genauigkeit zu erhöhen (Gandhi et al., 2023; Kiraly et al., 2024). Problematisch ist jedoch, dass unterschiedliche KI-Systeme nur schwer miteinander vergleichbar sind. In der Praxis zeigen sich mitunter deutliche Abweichungen in der Erkennungs- und Klassifikationsleistung, was Unsicherheiten bei der Auswahl der passenden Anwendung schafft (vgl. Yasaka & Abe, 2018; Kondrashova et al., 2024). Kondrashova et al. (2024) konnten beim Vergleich zweier Softwaresysteme relevante Abweichungen sowohl bei der Volumetrie als auch bei der Detektionsleistung aufzeigen. Chouffani et al. (2024) beschreiben zudem, wie unterschiedlich der Umfang der klinischen Erprobung trotz FDA-Zulassung ausfallen kann, und schlagen vereinheitlichte Prüfverfahren und Dokumentationsstandards für die Kennzeichnung vor.

Diese fehlende Vergleichbarkeit und variable Dokumentation unterstreichen die Bedeutung einer systematischen Qualitätssicherung: Zu den Hauptanforderungen gehören die Definition relevanter Kennzahlen – beispielsweise Sensitivität, Spezifität und Volumetrie-Genauigkeit – sowie die Etablierung standardisierter Testprotokolle. Dabei spielen Datensätze mit verlässlichem Goldstandard eine zentrale Rolle, um KI-Modelle objektiv und reproduzierbar beurteilen zu können. Solche Referenzdatensätze gewinnen durch die fortschreitende Integration von KI-Tools in die klinische Praxis zusätzlich an Bedeutung. Eine transparente Qualitätskontrolle wird für die Anwenderinnen und Anwender immer wichtiger (vgl. Jacobs et al., 2016; Allen & Dreyer, 2019).

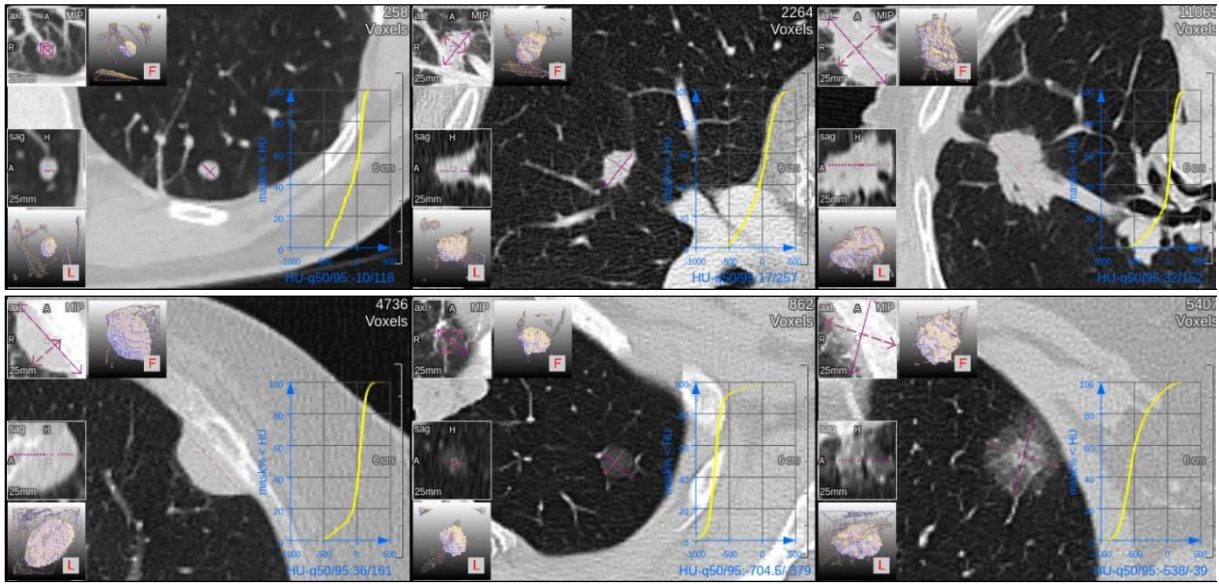

**Abb. 1:** Beispiele in-vivo erfasster Lungenläsionen, **obere Reihe:** Kleiner Rundherd Lung-RADS 3 (V=158 mm³, N=258); solider Tumor Lung-RADS 4A (V=693 mm³, N=2.264); solider Tumor Lung-RADS 4B (V=4.322 mm³, N=11.065); **untere Reihe:** Solider, pleuraständiger Tumor Lung-RADS 4B (V=4.517 mm³, N=4.736); Milchglastrübung Lung-RADS 2 (V= 580 mm³, N=862); teilsolider Tumor Lung-RADS 4B (V=4.364 mm³, N=5.407). **Erläuterungen:** In den Bildern sind jeweils zusätzlich eingeblendet (Overlays) eine axiale MIP und 3D-Darstellung der Tumorumgebung (oben), eine sagittal rekonstruierte Schicht und 3D-Darstellung (links) sowie das kumulierte HU-Histogramm unter der Segmentierungsmaske (rechts). Die angegebenen Volumina (V) und Anzahl der Voxel (N) beziehen sich ebenfalls auf diese Segmentierungsmaske. **Bildquelle:** Alle sechs Läsionen sind dem öffentlich frei verfügbaren Deep Lesion-Datensatz entnommen (US-NIH, veröffentlicht 2018).

Auch international lässt sich ein klarer Trend zur Etablierung standardisierter Prüfverfahren beobachten. In den USA arbeiten beispielsweise das Data Science Institute des American College of Radiology an „Certify AI" (Allen & Dreyer, 2019) und die Mayo Clinic an der KI-Plattform „Validate" – beides Netzwerkinitiativen, die einheitliche Methoden zur Bewertung von KI-Lösungen vorantreiben sollen. Ähnliche Projekte finden sich in Europa und Asien. So hat die aus den Niederlanden koordinierte Initiative „grand-challenge.org" mittlerweile weltweite Bedeutung erreicht, indem sie sich mit hunderten erfolgreich durchgeführten öffentlichen Wettbewerben der vergleichenden Wissenschaft in der KI für die medizinische Bildgebung verschrieben hat. Insgesamt steht nicht nur die punktuelle Leistungsbewertung im Vordergrund, sondern auch die Frage, wie sich eine kontinuierliche Validierung nach jeder Software-Aktualisierung realisieren lässt. In Deutschland betonen Fachgesellschaften wie die Deutsche Röntgengesellschaft (DRG) die Notwendigkeit nationaler Lösungen, die in internationale Netzwerke eingebunden sind (Vogel-Claussen et al., 2024).

Für die radiologische Praxis impliziert dies, dass künftig verstärkt auf normierte KI-Benchmarks und Referenzdatensätze zurückgegriffen werden soll, die realistische Szenarien in Screening oder Klinik abbilden und gleichzeitig exakte Messungen von Parametern, wie im Fall des Lungenkrebs-Screenings der Läsionsgröße, ermöglichen. Von zentraler Bedeutung ist daher die Definition eines geeigneten Goldstandards. Ziel dieser Arbeit ist es deshalb, eine Strategie aufzuzeigen, wie Ergebnisse automatisierter Verfahren überprüfbar und vergleichbar gemacht und verlustfrei in etablierte Befundungsprozesse integriert werden können. Letztlich liegt es im Interesse aller Akteure – von den Radiologinnen und Radiologen über Klinikträger bis hin zu den Herstellern, und insbesondere den am Screening Teilnehmenden sowie Patientinnen und Patienten –, einen hohen Qualitätsstandard zu etablieren, damit das positive Potential des Lungenkrebs-Screenings samt Computerunterstützung ausgeschöpft werden kann.

## 2. Regulatorik

In Deutschland und in der EU unterliegen KI-Anwendungen in der Medizin strengen Regularien. Sie werden als Medizinprodukte klassifiziert, wenn sie für diagnostische, therapeutische oder präventive Zwecke eingesetzt werden. Diese Einstufung beruht auf der Medizinprodukte-Verordnung (EU) 2017/745 (Medical Device Regulation, MDR), die in ihrer Fassung von Mai 2021 in Kraft ist und mehrere Risikoklassen unterscheidet.

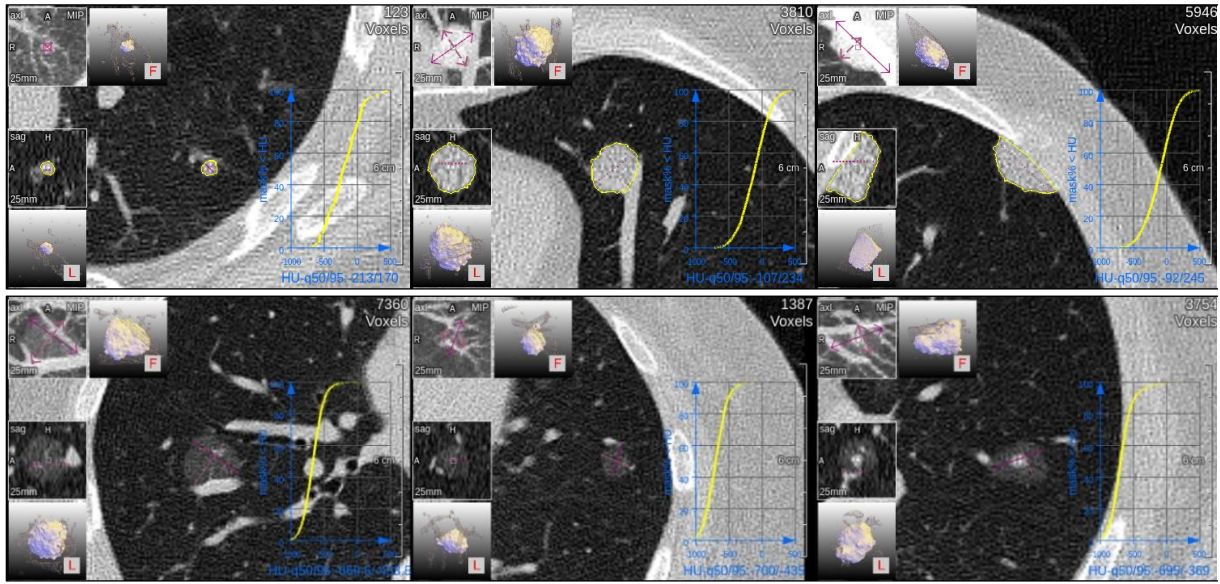

**Abb. 2:** Sechs simulierte solide Lungen-Läsionen (digitale Läsionsphantome) mit exakt bekannter Form und Größe, **obere Reihe:** Kleine typische Screening-Läsion (V=55 mm³, N=123); größere solide Läsion (V=1.713 mm³, N=3.810); pleuraständige Läsion (V=2.673 mm³, N=5.946); **untere Reihe:** Milchglastrübung (V=2.179 mm³, N=7.360, C=30% relativer Kontrast); subtilere Milchglastrübung (V=411 mm³, N=1.387, C=25%); größere teil-solide Läsion (V=1.688 mm³, N=3.754, C=25%). **Erläuterungen:** Die Overlays und Maßangaben sind identisch zu Abb. 1 gewählt. Anders als in Abb. 1 handelt es sich bei den Größenangaben jedoch um die wahren, durch die Simulation bekannten Maße. **Bildquelle:** Bei der Simulation wurde die exakt bekannte Form in jeweils ein unauffälliges Screening-Lungen-CT der HANSE-Studie (zur Verfügung gestellt durch J. Vogel-Claussen, MH Hannover, Aufnahme 2022) unter Beachtung der typischen Randunschärfe und Textur virtuell implantiert.

Eine KI-basierte Software, die Diagnose-Wahrscheinlichkeiten oder Behandlungsempfehlungen ausgibt, ist dabei in höhere Risikoklassen einzustufen (Klassen IIa, IIb oder III). Ob eine Software, die die befundende Person bei der Detektion unterstützt oder Herdbefunde volumetriert, bereits als hochriskant gilt, hängt von ihrem beabsichtigten Einsatz (engl. intended use) und von Art und Umfang der von ihr dargebotenen Informationen ab. Da bei der Lungenkrebsfrüherkennung die Größe der Herdbefunde das weitere Vorgehen direkt beeinflusst, ist in diesem Fall von einer höheren Risikoklasse auszugehen.

Für die Zulassung müssen Hersteller eine umfassende technische Dokumentation, Sicherheitstests, klinische Daten und Studien vorlegen, um die valide medizinische Funktion des KI-Systems zu belegen. Über eine benannte Stelle wird dann geprüft, ob diese Unterlagen den Anforderungen der MDR entsprechen. Bei positivem Ergebnis erhält das KI-Produkt eine CE-Kennzeichnung, die die Vermarktung im Europäischen Wirtschaftsraum ermöglicht.

Allerdings betont die Rechtsprechung (siehe LG Stendal, Aktenzeichen 31 O 50/08), dass das CE-Siegel nicht als Qualitätsgarantie, sondern primär als Nachweis der Konformität mit europäischen Richtlinien zu verstehen ist. Für die Vergabe eines CE-Siegels wird grundsätzlich die Leistung des entsprechenden Gerätes im Hinblick auf bestimmte – vom Hersteller behauptete – Leistungsparameter geprüft und nicht der Einsatz innerhalb eines bestimmten Prozesses, der unter Umständen andere Parameter erfordert.

Eine Studie ergab jüngst, dass von über 500 in den USA durch die Food and Drug Administration (FDA) zugelassenen KI-Anwendungen weniger als ein Drittel an prospektiv erhobenen Daten validiert wurden (Chouffani et al., 2024). Dies illustriert, dass die Charakteristika lernender Systeme – insbesondere in der Gesundheitsversorgung – noch unzureichend in bestehende Zulassungswege integriert sind. KI-Modelle verändern sich teils durch Daten-Updates oder algorithmische Anpassungen. Daher werden auch kontinuierliche Prüfungen nach der Zulassung notwendig (engl. post-market surveillance).

Ab 2026 verschärft der EU AI Act (EU-Verordnung 2024/1689) diese Anforderungen, indem KI-Systeme im Gesundheitswesen grundsätzlich als Hochrisiko-Anwendungen betrachtet werden. Hersteller müssen dann für all diese Systeme umfassende Qualitäts- und Risikomanagementprozesse etablieren und die Möglichkeit einer menschlichen Aufsicht sicherstellen. Zudem fordern die neuen Regelungen, dass Trainings-, Validierungs- und Testdaten relevant, repräsentativ und möglichst fehlerfrei sein müssen. Auch

Mechanismen, die Veränderungen in den Eingabedaten (engl. data drift) überwachen, werden als notwendig angesehen (Sahiner et al., 2023). Dadurch rückt die Idee eines systematischen, wiederkehrenden Qualitätstests in den Mittelpunkt, der sich an einem sorgfältig kuratierten Referenzdatensatz orientiert und reproduzierbare Evaluationen ermöglicht (Huo et al., 2013).

## 3. CT-Technik

Die Computertomographie (CT) ist die Methode der Wahl zur Lungenkrebs-Früherkennung. Aufgrund kontinuierlicher Entwicklungen in der Gerätetechnologie existieren derzeit in der klinischen Routine eine Vielzahl von Geräten, Techniken und Protokollen. Insbesondere in den letzten Jahren konnten nochmals signifikante Fortschritte erzielt werden, unter anderem durch die Einführung neuer Photon-Counting-Detektorsysteme (Symons et al., 2017) sowie durch die Entwicklung KI-gestützter Rekonstruktionsverfahren (Jiang et al., 2022).

Für die Lungenkrebs-Früherkennung wird gefordert, adaptive Rekonstruktionsverfahren (iterativ oder KI-basiert) anzuwenden. Diese können das Bildrauschen reduzieren und die Bildqualität insgesamt erhöhen, wobei die Zusammenhänge für eine optimale Parameterwahl komplex sind (Greffier et al., 2022). Darüber hinaus sind standardisierte Niedrigdosis-Protokolle erforderlich, die eine an die Körperstatur angepasste Röhrenspannung von typischerweise 100 bis 120 kV und einen automatisch angepassten Röhrenstrom verwenden, um eine ausreichende Bildqualität für die Detektion kleiner Lungenrundherde zu gewährleisten. Der zugehörige Volumen-Computertomografie-Dosisindex ($CTDI_{vol}$) liegt dabei typischerweise unterhalb 1,3 mGy für eine Standardperson (Vogel-Claussen et al., 2024). Diese und weitere Anforderungen sind im Anhang zur Lungenkrebsfrüherkennungsverordnung des BMUV festgelegt und zielen darauf ab, die Strahlenexposition der teilnehmenden Personen zu minimieren und gleichzeitig die diagnostische Genauigkeit zu maximieren (BMUV 2024).

Die nach DIN EN IEC 61223-3-5 etablierten gerätebezogenen Abnahme- und Konstanzprüfungen werden aktuell als ausreichend angesehen und sollten weiterhin jährlich sowie jeweils nach wesentlichen Wartungstätigkeiten (z.B. Röhrentausch, Detektortausch etc.) durchgeführt werden. Für die Qualität der Volumenbestimmung können die niedrige Dosis und das dadurch erhöhte Rauschen eine größere Variabilität verursachen (Guedes Pinto et al., 2023).

Die historische Vielfalt an technischen Geräten und Konfigurationen stellt ein grundsätzliches Problem für KI-Modelle dar, da unterschiedliche Protokolle und Qualitätsniveaus zu sog. Domain-Shifts in den Bilddaten führen können. Hierbei wird die Komplexität der Zusammenhänge zwischen Dosis, Bildqualität und der Integration von KI-Systemen in die klinische Praxis deutlich. KI-Algorithmen, die ausschließlich auf Datensätzen von High-End-Scannern trainiert werden, könnten etwa in Einrichtungen mit älteren Systemen weniger genau sein. Der Zusammenhang zwischen CT-Dosis, gewähltem Bildrekonstruktionsverfahren und Bildqualität mit der resultierenden Genauigkeit von KI-Algorithmen ist komplex und erfordert eine eingehende Prüfung der Ergebnisqualität (Gupta et al., 2022).

Um die Kompatibilität von KI-Algorithmen mit verschiedenen CT-Systemen und gleichzeitig insgesamt eine hohe Ergebnisqualität zu sichern, ist eine definierte Mindestqualität essenziell. Dies sollte auch in die Dokumentation der Studien zur Validierung mit einfließen. Zudem sind Testobjekte (Phantome) hilfreich, um regelmäßig die Bildqualität und Messgenauigkeit zu kontrollieren (D'hondt et al., 2024). Bei der Einschätzung von Lungenrundherden, insbesondere bei der Beurteilung der Wachstumsraten, ist zudem ein konsistentes Scan-Protokoll zwischen aufeinanderfolgenden Untersuchungen bedeutend, sodass Veränderungen im Volumen tatsächlich detektiert und nicht durch Protokolländerungen überlagert werden. Zusätzliche Variabilität kann durch die Verwendung unterschiedlicher Softwaresysteme impliziert werden (Avila et al., 2023; QIBA, 2018).

Es kann davon ausgegangen werden, dass die Hersteller für den jeweiligen Gerätetyp und die im Einsatz befindlichen Softwareversionen geeignete Protokolle zur Verfügung stellen. Diese sollten eindeutig benannt und von den für das Gerät zuständigen Expertinnen und Experten aus Radiologie und Medizinphysik für die klinische Anwendung freigegeben werden. Zur Kontrolle einer konsistenten Nutzung der Scan-Protokolle wird zudem die Verwendung eines Dosismanagementsystems empfohlen.

Eine systematische Dokumentation der Protokolle und Dosiswerte hätte großen Nutzen für strukturierte wissenschaftliche Auswertungen ihres Einflusses auf die Ergebnisqualität und für die Identifikation weiterer

Optimierungspotentiale. Auf Basis solcher Auswertungen sollten die Empfehlungen zu geeigneten Parametern regelmäßig aktualisiert werden. Damit könnte nicht nur die technische Qualitätssicherung evaluiert werden, sondern es könnten auch verlässliche Angaben zur Detektierbarkeit von Rundherden in Abhängigkeit von deren Größe (Volumen in mm³) und Dichte (Schwächung in HU) formuliert werden, die im Screening erreicht werden können.

## 4. Qualitätsbeurteilung von KI-Modellen

Die Qualitätsbeurteilung von KI-Systemen zur Lungenkrebs-Früherkennung erfordert eine Anzahl quantitativer Maßzahlen und Testkriterien, um Objektivität und Aussagekraft zu gewährleisten (Hadjiiski et al, 2023). Ein wichtiger Aspekt bei der Evaluation der Detektionsleistung ist die sog. Standalone-Performance. Darunter versteht man die erzielbare Genauigkeit eines KI-Systems in einem vollautomatischen Modus, also ohne Eingriffe oder Korrekturen durch menschliche Expertinnen und Experten. Zwar werden KI-Anwendungen in der Routine häufig in einem interaktiven Szenario verwendet werden, bei dem Radiologinnen oder Radiologen die Softwareergebnisse nachbearbeiten oder selektiv verwenden bzw. bewerten, doch ermöglicht erst die Analyse im Standalone-Modus eine objektive, vergleichbare Bewertung der rein algorithmischen Detektions- und Klassifikationsfähigkeit. Diese systematische Evaluierung ist in der Lage aufzuzeigen, welches Leistungsniveau ein KI-System ohne zusätzliche Optimierung bzw. Korrekturen durch den Anwender tatsächlich erreicht, und liefert damit wertvolle Vergleichsmaße unterschiedlicher KI-Modelle (Sahiner, 2012).

Zur quantitativen Bewertung der Detektionsgüte werden etablierte Metriken wie Sensitivität und Spezifität herangezogen. Diese Kennzahlen ermöglichen eine umfassende Einschätzung der Leistungsfähigkeit eines Detektionsverfahrens, da sie sowohl falsch-positive Befunde als auch nicht entdeckte Läsionen berücksichtigen. Weitere wichtige Kennzahlen zur Charakterisierung eines Detektionssystems sind der sog. positive prädiktive Wert sowie der negative prädiktive Wert (Abk. PPV, NPV), die jeweils angeben, zu welchem Prozentsatz ein positives bzw. negatives Testergebnis korrekt ist.

---

### *Strukturierte Befundung der LDCT zur Lungenkrebs-Früherkennung mit geeigneter Befundungssoftware*

- Die Strukturierte Befundung soll einheitlich nach der modifizierten Lung CT Screening Reporting & Data System Klassifikation (Lung-RADS v2022) mit Integration der Volumenverdopplungszeit durchgeführt werden. Dabei kommt entsprechend der BMUV-Verordnung eine qualitätsgesicherte Software zur Befundunterstützung zum Einsatz, für die folgende Leistungsmerkmale gelten:
    a) Detektion der Lungenrundherde
    b) Volumetrie der Lungenrundherde
    c) Berechnung der Volumenverdopplungszeit (VDT) im Vergleich mit dem Vorbefund
    d) Beschreibung der Lungenrundherde in folgenden Kategorien (*):
       solide; teilsolide; Milchglas; Verkalkungen; juxtapleural
    e) Automatische Kategorisierung der Lung-RADS-Klasse (*)
    f) Berechnung des Malignitätsrisikos des Lungenrundherdes (*)

- Aus den notwendigen Merkmalen a–c leitet sich der dringende Bedarf der Qualitätsanalyse insbesondere für die Genauigkeit und Zuverlässigkeit der Detektion und Volumetrie ab. (*) Die Merkmale d–e sind bislang noch nicht in den Anforderungen der Verordnung enthalten.

- Wünschenswert wäre zudem die automatische Erstellung einer Liste zur Zweitbefundung mit der Möglichkeit der regionalen und nationalen Vernetzung. Auch sollten zur Fehlerminimierung identische Softwaresysteme bei der Erst- und Zweitbefundung eines Falles genutzt werden.

- Da es durch unterschiedliche Software z. B. zu Unterschieden in der Volumetrie und in der vorgeschlagenen Kategorisierung kommen kann, spricht sich die DRG für vergleichende und qualitätsbestimmende Studien aus. Aus deren Ergebnissen können Empfehlungen abgeleitet werden für Kriterien für den Einsatz der Software in einem zukünftigen strukturierten nationalen Lungenkrebs-Screening-Programm.

Neben der reinen Detektion spielt die strukturierte Charakterisierung der erkannten Läsionen eine wichtige Rolle. Insbesondere sind Volumen und Dichte von hoher klinischer Relevanz, da sie maßgeblich in Leitlinien wie Lung-RADS oder nationalen Empfehlungen zur Beurteilung von Lungenrundherden berücksichtigt werden. Die Beurteilung der Volumetrie-Genauigkeit bei in-vivo erhobenen Datensätzen erfordert eine sorgfältige Auswahl der Referenzsegmentierung, da diese maßgeblich die Aussagekraft des Vergleichs beeinflusst. Besonders bei kleinen Volumina ist der Einfluss des Partialvolumeneffekts erheblich, was zu einer erhöhten Inter-Reader-Variabilität führt (Lassen et al., 2015), die sich beispielsweise in niedrigen Dice-Scores zeigt, sowie systematisch zu einer Volumenüberschätzung (Rexilius et al., 2005). Diese Fehlerquellen sind bei der Gegenüberstellung eines Modells mit der Referenzsegmentierung zu berücksichtigen, um eine realistische Einschätzung der Modellgenauigkeit zu gewährleisten. Eine robuste Validierung muss sich daher auf standardisierte Referenzdatensätze und eine methodische Analyse der Variabilität zwischen unterschiedlichen Expertinnen und Experten beziehen.

Da die exakten, tatsächlichen Läsions-Volumina bei klinisch erhobenen Daten grundsätzlich unbekannt sind, muss die Evaluierung der absoluten Volumetrie-Genauigkeit eines KI-Systems ergänzend oder alternativ durch Kontrollen mithilfe physischer oder simulierter Phantom-Datensätze (s. Abb. 2) überprüft werden. Die Simulation von Läsionen für Volumetrie-Studien wurde initial für Gehirnläsionen durchgeführt (Rexilius et al., 2005). Die absolute Genauigkeit der Volumenmessung spielt insbesondere bei Szenarien eine Rolle, bei denen unterschiedliche KI-Systeme innerhalb einer Screening- oder Versorgungsregion zum Einsatz kommen würden. Darüber hinaus bietet es sich an, in Studien zur Nachverfolgung (Follow-Up-Scans) die räumliche Zuordnung (Bildregistrierung) und relative volumetrische Messgenauigkeit wiederholter Untersuchungen (Reproduzierbarkeit) zu bewerten (de Hoop et al., 2009).

Manche modernen KI-Systeme gehen über die bloße Detektion, Volumetrie und Dichtebeurteilung hinaus und geben Malignitätsabschätzungen oder verweisen auf Leitlinien für das weitere Prozedere. Diese Modelle zeigen aber eine erhebliche Variabilität, und die Genauigkeit hängt entscheidend von der klinischen Anwendung und der Auswahl der Trainingsdaten ab (Heideman et al., 2024; Li et al., 2025). In solchen Fällen ist, passend zu der oben genannten höheren Risikoklasse, eine noch umfassendere Qualitätsbeurteilung notwendig, die auch die Genauigkeit der Klassifikation oder die Korrektheit der Handlungsempfehlungen einschließt.

Wichtig ist, dass die verwendeten Metriken eine Relevanz für Screening oder Klinik haben und sich in den medizinischen Kontext übertragen lassen. Ein nicht hinreichend überprüftes KI-System kann etwa in einem experimentellen Setting hohe Übereinstimmungswerte bei der Segmentierung erzielen und dennoch an für die klinische Entscheidungsfindung entscheidenden Stellen fehlerhaft sein.

Um den praktischen Nutzen eines KI-gestützten Systems zu gewährleisten, ist zudem die reibungslose Integration in den Workflow entscheidend. Da sich die Arbeitsabläufe und Anforderungen zwischen radiologischen Abteilungen unterscheiden, muss die Eignung eines KI-Systems individuell geprüft werden. Dafür sollten relevante Eckdaten systematisch bereitgestellt werden, um eine fundierte Bewertung und ggf. Anpassung an die spezifischen Bedürfnisse der jeweiligen Einrichtung zu ermöglichen.

## 5. Referenzdatensätze

Der Aufbau und die Pflege geeigneter Referenzdaten sind zentrale Bestandteile der Qualitätssicherung. Ein solcher Datensatz soll in repräsentativer Weise verschiedene Gerätetypen, CT-Protokolle und Personengruppen aus mehreren Standorten abdecken (Armato et al., 2011). Er muss strukturiert sein, damit für jedes Bild die relevanten Metadaten (z. B. Scanparameter) und Annotationen (z. B. Markierungen von Lungenrundherden, Volumensegmentierungen, klinischer Verlauf) vorliegen. Eine Rolle spielen dabei Expertenannotationen, die i. d. R. den Goldstandard definieren. Um die Interrater-Variabilität des Goldstandards und Ungenauigkeiten durch die Bildrekonstruktion sowie -darstellung zu erfassen, sollten im Idealfall mehrere Radiologinnen und Radiologen unabhängig voneinander segmentieren oder klassifizieren.

Ein Referenzdatensatz für Screening-spezifische Fragestellungen muss nicht die gesamte Bandbreite der Pneumologie abdecken, aber unbedingt eine repräsentative Anzahl jener Fälle einschließen, die in diesem Anwendungsbereich besonders relevant sind. Dazu gehören verschiedene Befundkonstellationen (z. B. solider oder semi-solider Rundherd und Milchglastrübungen, vgl. Abb. 1 & 2), Varianten der Bildqualität (z. B. vorliegende Artefakte oder schwache Kontrastverhältnisse), sowie ggf. vorhandene Komorbiditäten und weitere Besonderheiten (z.B. besondere Anatomien nach Lappenresektion). Um die Volumenmessung zu evaluieren, werden daneben physische Phantome und softwarebasierte Simulationen virtueller Läsionen eingesetzt (s. o., Abb. 2), von denen die exakten Volumina bekannt und in der Referenzdatenbank hinterlegt sind.

In ähnlicher Weise, wie über Phantom-Datensätze Hinweise zur absoluten Genauigkeit der Volumetrie ermittelt werden können, wäre es auch für die Detektions-Leistung wünschenswert, einen absoluten Referenzpunkt zu generieren. Die BMUV-Verordnung zielt bezüglich der Detektion auf „kontroll- und abklärungsbedürftige Befunde". Deren Definition ist jedoch untersucherabhängig und nicht perfekt reproduzierbar, da sie von individuellen Erfahrungshintergründen und Befundungsroutinen der beteiligten Radiologinnen und Radiologen abhängt.

Wir schlagen vor, die Detektions-Leistung eines KI-Systems nicht allein auf visuell identifizierte Befunde zu beziehen, sondern zusätzlich auf einen endpunktbezogenen Goldstandard, der eine eindeutigere Vergleichsmetrik bietet. So sollte die Standalone-Performance hinsichtlich Sensitivität und Spezifität perspektivisch auch anhand einer großen Anzahl retrospektiv gesicherter Fälle erfolgen. Bei diesen liegt eine entsprechende Verlaufsbeobachtung bzw. histopathologische Sicherung vor, sodass der jeweilige Status als „maligne" oder „benigne" bzw. „ohne Befund" zuverlässig bekannt ist.

Ein weiterer Aspekt ist die laufende Aktualisierung. Mit der Einführung neuer CT-Technologien wie Photon-Counting (Symons et al., 2017; Rajendran et al., 2022) und der Weiterentwicklung von Bildrekonstruktionsverfahren müssen entsprechende Datensätze nachgezogen werden (vgl. Jiang et al., 2022). Andernfalls könnten veraltete Bildqualitäten dominieren, was die Aussagekraft einer Prüfung mindert.

## 6. Bereitstellung

Die praktische Umsetzung solcher Prüfverfahren bzw. Zertifizierungen ist komplex. Ein wichtiger Diskussionspunkt ist, wie Hersteller bzw. deren Produkte Zugang zum Benchmarking erhalten und gleichzeitig verhindert wird, dass die Testfälle in falsche Hände geraten oder sogar gezielt als Entwicklungsdaten mit dem Ziel einer unlauteren Verbesserung der Testergebnisse genutzt werden.

Es besteht also grundsätzlich die Notwendigkeit, die Daten streng zu schützen und nur berechtigten Stellen zum Zwecke der Prüfung zugänglich zu machen. Von den Herstellern wäre dann zu verlangen, dass sie etwa eine Container-basierte Version ihrer Software bei diesen Stellen ausführen lassen (vgl. Boettiger, 2014). Container bezeichnen ein softwaretechnisches Konstrukt, bei dem KI-Lösungen unter standardisierten Bedingungen auf einem anderen Server ausgeführt werden können. Für die Container-basierte Prüfung ist vorab eine Testumgebung zur Verfügung zu stellen, mit der die Integrations-Schnittstelle erprobt werden kann.

Um das Risiko einer Abstimmung der Testdaten zu minimieren und die Datenintegrität zu wahren, empfiehlt die Food and Drug Administration (FDA, 2022) zudem die Einrichtung eines sog. Prüfpfades. Dieser beinhaltet (a) eine zufällige Auswahl der Testdaten aus einer größeren Testdatenbank, (b) die Festlegung einer Grenze, wie oft ein Fall für die Bewertung verwendet werden kann, (c) die Kontrolle des Datenzugriffs, um sicherzustellen, dass nur zusammengefasste Leistungsergebnisse die Testumgebung verlassen, sowie (d) das Führen eines Datenzugriffsprotokolls, einschließlich einer Aufzeichnung darüber, wer auf die Daten zugegriffen hat, der Testbedingungen und der standardisiert zusammengefassten Leistungsergebnisse (z. B. Sensitivität, Spezifität, Volumengenauigkeit etc.).

Darüber hinaus ist es wünschenswert, dass Teile des Datensatzes – für Forschungszwecke und Vorab-Tests sowie insgesamt zur Stärkung der Transparenz – öffentlich verfügbar sind (Prior et al., 2020; Jacobs et al., 2021). Durch eine sorgfältige Balance zwischen Offenheit und Geheimhaltung, insbesondere durch eine strikte Trennung von Referenz- bzw. Testdaten auf der einen und Forschungsdaten auf der anderen Seite, kann die Objektivität der Prüfung abgesichert werden.

Wie erwähnt, stellt eine rein automatisierte Testung jedoch nicht in jedem Fall das reale Anwendungsszenario dar, weil manche KI-Systeme die Interaktion mit Radiologinnen oder Radiologen erfordern. Eine mögliche Lösung ist ein gestaffeltes Verfahren: Zuerst wird ein automatischer Teiltest ohne Interaktionen durchgeführt, der zu einer vorläufigen Bewertung führt, und im weiteren Verlauf wird eine Interaktionsphase simuliert oder realisiert, die das Bild vervollständigt. Hierfür wurden ergänzende Plattformen entwickelt, die eine strukturierte Bewertung anhand von Stichproben im klinischen Einsatz ermöglichen (Nomura et al., 2020).

Aufbau und Pflege eines ausreichend großen, kontinuierlich aktualisierten Datensatzes sind aufwändig, weshalb zwischen Reliabilität der Testung und ökonomischer Durchführbarkeit abgewogen werden muss. Daher könnte ein Mischmodell sinnvoll sein, bei dem staatliche Fördermittel und Herstellerbeiträge die Kosten gemeinsam decken. Eine gewisse Gebühr für jeden Testlauf wäre denkbar, um die operativen Kosten zu decken. Auf diese Weise ließe sich sicherstellen, dass nur ernsthaft interessierte Anbieter den Dienst nutzen, während gleichzeitig ein nachhaltiger Betrieb eines Prüfverfahrens gewährleistet werden kann.

## 7. Outreach

Aus Sicht der DRG eröffnet die Qualitätssicherung von KI-Modellen für die Lungenkrebs-Früherkennung weitreichende Perspektiven für die Zukunft der radiologischen Diagnostik. Als eine der treibenden Kräfte im deutschsprachigen Raum trägt die DRG maßgeblich dazu bei, Standards zu setzen, Leitlinien weiterzuentwickeln und den technischen Fortschritt verantwortungsbewusst in den klinischen Alltag zu integrieren, auch als Vorreiter für andere Fachgesellschaften (s.a. ESR, 2019). Die Anforderungen an Referenzdatensätze, Metriken und Genehmigungs- oder Zulassungsprozesse dienen nicht nur der Lungenkrebs-Früherkennung, sondern besitzen auch Modellcharakter für andere Anwendungsgebiete der medizinischen Bildgebung.

Ein wesentliches Transferpotenzial besteht darin, Verfahren und Kriterien der Qualitätssicherung auf weitere bildgebungsbasierte KI-Anwendungen zu übertragen, etwa in der Mammadiagnostik, in der Prostatakrebsfrüherkennung, in der Neuroradiologie oder bei muskuloskelettalen Fragestellungen (vgl. Benjamens et al., 2020). Dies erfordert eine enge Abstimmung in und zwischen den DRG-Arbeitsgruppen, Fachgesellschaften, medizinischen oder berufsgruppenbezogenen Organisationen und Gremien. Die Zusammenarbeit mit der Medizinphysik, der medizinischen Informatik, den Ingenieurwissenschaften und regulatorischen Stellen ist entscheidend für die gemeinschaftliche Entwicklung und Evaluierung belastbarer Testprotokolle.

Standardisierte und vergleichbare Maßzahlen für KI-Modelle verschaffen allen Anwenderinnen und Anwendern eine solide Entscheidungsgrundlage. Ein strukturiertes Prüf- und Bewertungsverfahren kann die Auswahl zwischen unterschiedlichen kommerziellen Systemen verbessern, indem es die bislang oft abstrakten Leistungsparameter einer KI mit direktem Praxisbezug greifbar und vergleichbar macht (Park et al., 2023). Bei der Gesamtbewertung darf der Fokus allerdings nicht allein auf den technischen Kennzahlen liegen, sondern es sind auch die potenziellen Vorteile im Ablauf in den radiologischen Zentren, etwa verbesserte Interaktionsmöglichkeiten, zu berücksichtigen.

Die rasante Entwicklung der KI-Technologien zur Bilderkennung unterstreicht die Rolle der Radiologie als eines der innovativsten Fächer der Medizin (Gandhi et al., 2023). Die DRG begleitet diese Entwicklung auf mehreren Ebenen, um die vielseitigen Ansprüche hinsichtlich Ideenförderung, Digitalisierung, Patientensicherheit und Versorgungsqualität zu identifizieren und zu unterstützen.

Klare Maßstäbe für den Einsatz von KI – insbesondere in der Früherkennung von Lungenkrebs – stärken die Verlässlichkeit von und damit auch das Vertrauen in neue Technologien. Dies ist nicht nur für Universitätsklinika relevant, sondern insbesondere auch für kleinere Häuser und niedergelassene Radiologie-Praxen, die von strukturierter Befundung und klar definierten Qualitätskriterien profitieren können. Nicht zuletzt soll dies Positionspapier eine Hilfestellung für Aufsichtsbehörden und regulatorische Stellen bei der Umsetzung von abstrakten Gesetzestexten in praxistaugliche Prüfprotokolle sein.

## Literaturverzeichnis